\documentclass[12pt]{article}

\usepackage{amsmath}
\usepackage{amssymb}
\usepackage{amsthm}
\usepackage[toc]{appendix}
\usepackage{caption}
\usepackage{chngcntr}
\usepackage{xcolor}
\usepackage{enumerate}
\usepackage[Conny]{fncychap}
\usepackage{float}
\usepackage[top=2cm, bottom=2cm, left=2cm, right=2cm]{geometry}
\usepackage{graphicx}
\usepackage[linktoc=all]{hyperref}
\usepackage{mathrsfs}
\usepackage{marvosym}
\usepackage{pifont}
\usepackage{thmtools}
\usepackage{tikz-cd}
\usepackage{subcaption}

\usepackage{mdframed}

\usepackage{titlesec}
\usepackage[object=vectorian]{pgfornament}
\usepackage{cleveref}
\usepackage{appendix}
\titleformat{\section}
  {\normalfont\Large\bfseries\MakeUppercase}{\thesection}{1em}{}

\tikzcdset{row sep/Large/.initial=5em}
\tikzcdset{column sep/Large/.initial=5em}
\definecolor{theorem_color}{rgb}{0.35,0.0,0}
\definecolor{VividBurgundy}{RGB}{159,29,53}
\definecolor{Burgundy1}{RGB}{128,0,32}
\definecolor{darkbrown}{rgb}{0.4, 0.26, 0.13}
\newcommand{\OldYellow}{brown!5!white!95}

\newcommand{\themecolor}{brown!70!black}

\declaretheoremstyle[%
  spaceabove=4pt,%
  spacebelow=1pt,%
  headfont=\normalfont\bfseries,%
  notefont=\bfseries\itshape, notebraces={}{},%
  bodyfont=\normalfont\slshape,%\itshape
  postheadspace=1em,%
	mdframed={
		leftline=true,           % Enable the left line (ribbon)
		rightline=false,
		topline=false,
		bottomline=false,
		linecolor=\themecolor,          % Ribbon color
		linewidth=3mm,           % Ribbon width
		skipabove=10pt,          % Space above the environment
		innerleftmargin=10pt,    % Margin between the ribbon and text
		innerrightmargin=10pt,   % Margin on the right side of text
		innertopmargin=0pt,
		innerbottommargin=5pt
	}
]{theoremstyle} 

\declaretheoremstyle[%
  spaceabove=8pt,%
  spacebelow=8pt,%
  headfont=\normalfont\bfseries,%
  notefont=\bfseries\upshape, notebraces={}{},%
  bodyfont=\normalfont,%\itshape
  postheadspace=1em,%
]{problemstyle}

\declaretheoremstyle[%
  spaceabove=8pt,%
  spacebelow=8pt,%
  headfont=\normalfont\bfseries,%
  bodyfont=\normalfont,%\slshape,%\itshape
]{examplestyle}

\declaretheoremstyle[%
  spaceabove=0em,%
  spacebelow=1em%10pt,%
  headfont=\normalfont\itshape,%
  notefont=\bfseries, notebraces={}{},%
  postheadspace=1em,%
  bodyfont=\normalfont\upshape\color{black!90!gray},%
  qed=\qedsymbol,%
]{proofstyle} 

\declaretheoremstyle[%
  spaceabove=-6pt,%
  spacebelow=6pt,%
  headfont=\normalfont\bfseries,%
  postheadspace=1em,%
  bodyfont=\normalfont\upshape,%
]{remarkstyle} 

\declaretheoremstyle[%
  spaceabove=8pt,%
  spacebelow=8pt,%
  headfont=\normalfont\bfseries,%
  bodyfont=\normalfont\itshape,%
  postheadspace=1em,%
]{conjecturestyle}

\declaretheoremstyle[%
  spaceabove=8pt,%
  spacebelow=8pt,%
  headfont=\normalfont\bfseries,%
  bodyfont=\normalfont\itshape,%
  postheadspace=1em,%
]{questionstyle}

\declaretheoremstyle[%
  spaceabove=8pt,%
  spacebelow=12pt,%
  headfont=\normalfont\bfseries,%
  bodyfont=\normalfont,%\slshape,%\itshape
  postheadspace=1em,%
  notefont=\bfseries,
  notebraces={}{},
  headpunct={},
  mdframed={
	backgroundcolor=\OldYellow,
    linecolor=black,
    linewidth=1pt,
	innertopmargin=0.5em,
	innerbottommargin=1em,
    innerleftmargin=1em,
    innerrightmargin=1em,
    skipabove=2pt,
    skipbelow=2pt
  }
]{definitionstyle} 

\declaretheoremstyle[%
  spaceabove=4pt,%
  spacebelow=8pt,,%
  headfont=\normalfont\bfseries,%
  bodyfont=\small,%\normalfont,%\slshape,%\itshape
  postheadspace=1em,%
 	mdframed={
 	backgroundcolor=\OldYellow,
 	linecolor=\OldYellow,
     linewidth=1pt,
     innerleftmargin=1em,
     innerrightmargin=1em,
 	innertopmargin=2pt,
     skipabove=2pt,
     skipbelow=2pt,
  }
]{exercisestyle} 

\declaretheoremstyle[%
  spaceabove=1em,%
  spacebelow=1em,%
  headfont=\normalfont\bfseries,%
  bodyfont=\normalfont,%\slshape,%\itshape
  postheadspace=1em,%
  notefont=\bfseries,
  notebraces={}{},
  headpunct={},
  qed=$\diamondsuit$%
]{pointstyle}

\declaretheoremstyle[%
  spaceabove=8pt,%
  spacebelow=8pt,%
  headfont=\normalfont\bfseries,%
  bodyfont=\normalfont,%\slshape,%\itshape
  postheadspace=1em,%
  notefont=\bfseries,
  notebraces={}{},
  headpunct={},
	mdframed={
		leftline=true,           % Enable the left line (ribbon)
		rightline=false,
		topline=false,
		bottomline=false,
		linecolor=black,          % Ribbon color
		linewidth=1mm,           % Ribbon width
		skipabove=5pt,          % Space above the environment
		innerleftmargin=10pt,    % Margin between the ribbon and text
		innerrightmargin=10pt,   % Margin on the right side of text
		innertopmargin=-6pt,
		innerbottommargin=0pt
	}
]{illustrationstyle}

\declaretheorem[name={Theorem},style=theoremstyle,numberwithin=section]{theorem}
\declaretheorem[name={Lemma},style=theoremstyle,sibling=theorem]{lemma}
\declaretheorem[name={Proposition},style=theoremstyle,sibling=theorem]{proposition}
\declaretheorem[name={Corollary},style=theoremstyle,sibling=theorem]{corollary}

\declaretheorem[name={\textit{Proof}},style=proofstyle,unnumbered,]{prf}

\declaretheorem[name={\S } Definition,style=definitionstyle,unnumbered]{definition}

\declaretheorem[name=\ding{45},style=pointstyle,sibling=theorem]{point}

\numberwithin{equation}{section}

\definecolor{maroon}{rgb}{0.5, 0.0, 0.0} % Define the maroon color
\definecolor{gold}{rgb}{0.85, 0.65, 0.13} % Define the gold color
\hypersetup{
    colorlinks=false,
    linkbordercolor=gold,  % RGB values for brown
    citebordercolor=brown,    % RGB values for green
    urlbordercolor=maroon        % RGB values for blue
}

\renewcommand{\qedsymbol}{$\blacksquare$}

\newlength{\wideitemsep}
\let\olditem\item
\renewcommand{\item}{\setlength{\itemsep}{\wideitemsep}\olditem}

\newcommand{\ds}{\displaystyle}

\newcommand{\mc}{\mathcal}

\newcommand{\qqra}{\qquad\Rightarrow\qquad}

\newcommand{\lrb}[1]{\left[#1\right]}

\newcommand{\lrp}[1]{\left(#1\right)}

\newcommand{\set}[1]{\{#1\}}
\newcommand{\suppressthis}[1]{}

\newcommand{\F}{\mathbb F}
\newcommand{\E}{\mathbb E}

\newcommand{\Prob}{\mathbb P}

\newcommand{\R}{\mathbb R}

\AtBeginEnvironment{subappendices}{%
\counterwithin{figure}{section}
\counterwithin{table}{section}
}

\AtEndEnvironment{subappendices}{%
\counterwithout{figure}{section}
\counterwithout{table}{section}
}

\title{Vector Trifference}

\author{
  Siddharth Bhandari\thanks{Toyota Technological Institute at Chicago (Work done prior to author joining Amazon). Email: \texttt{siddharth@ttic.edu}}
  \and
  Abhishek Khetan\thanks{Ashoka University, Dept. of Mathematics. Email: \texttt{abhishek.khetan@ashoka.edu}}
}

\begin{document}

\maketitle
\pagenumbering{arabic}

\begin{abstract}
We investigate a geometric generalization of \emph{trifference}, a concept introduced by Elias~\cite{Elias1988} in the study of zero-error channel capacity. In the discrete setting, a code $\mathcal{C} \subseteq \{0,1,2\}^n$ is \emph{trifferent} if for any three distinct codewords $x,y,z \in \mathcal{C}$, there exists a coordinate $i \in [n]$ where $x_i, y_i, z_i$ are mutually distinct. Determining the maximum size of such codes remains a central open problem; the classical upper bound $|\mathcal{C}| \leq 2 \cdot (3/2)^n$ from \cite{Elias1988}, proved via a simple `pruning' argument, has resisted significant improvement.

Motivated by the search for new techniques, and in line with the \emph{vectorial extensions} of other classical combinatorial notions, we introduce the concept of \emph{vector trifferent} codes. 
Consider $\mathcal{C} \subseteq (S^2)^n$, where the alphabet is the unit sphere $S^2 = \{ v \in \mathbb{R}^3 : \|v\| = 1 \}$. We say the code $\mc C$ is vector trifferent if for any three distinct $x,y,z \in \mathcal{C}$, there is an index $i$ where the vectors $x_i, y_i, z_i$ are mutually orthogonal. 
A direct reduction of the vectorial problem to the discrete setting appears
infeasible, making it difficult to replicate Elias’s pruning argument.
Nevertheless, we develop a new method to establish the following upper bound for the size of a vector trifferent code:
\[
|\mathcal{C}| \leq (\sqrt{2} + o(1)) \cdot (3/2)^n.
\]
Interestingly, our approach, when modified and adapted back to the discrete setting, yields a polynomial improvement to Elias's bound: $|\mathcal{C}|\lesssim n^{-1/4}\cdot (3/2)^n$,
This improvement arises from a technique that parallels, but is not identical to, the method of~\cite{bhandari_khetan}. However, it still falls short of the sharper $n^{-2/5}$ factor obtained there. We also generalize the concept of vector trifferent codes to richer alphabets and prove a vectorial version of the Fredman-Komlós theorem~\cite{FredmanK1984-perfect} for general $k$-separating codes.
\end{abstract}

\newpage
\section{Introduction}

In this work we study a geometrical generalization of the concept of `trifference' define below.
\vspace{1em}
\begin{definition}[(Trifferent Codes)]
	\label{defn:trifference}
    For an integer $n>0$ we say three distinct strings $x,y$ and $z$ all belonging to $\{0,1,2\}^n$ are \emph{trifferent} if there exists a coordinate $i \in \{1, \dots, n\}$ where the entries $x_i, y_i, z_i$ are mutually distinct. 
	A \textit{trifferent code} of block length $n$ is a subset $\mathcal{C} \subseteq \{0,1,2\}^n$ such that any three distinct codewords $x, y, z \in \mathcal{C}$ are trifferent.
\end{definition}

The problem of finding the maximum possible size of a trifferent code of block length $n$ was first introduced by Elias \cite{Elias1988} and has occupied an important position at the intersection of zero-error information theory and combinatorics of hash functions. 
Upper bounding the maximum possible size of such a code, and understanding this question has drawn considerable attention. (See for instance the $2014$ Shannon Lecture: \href{https://www.itsoc.org/video/isit-2014-mathematics-distinguishable-difference}{`On The Mathematics of Distinguishable Difference'} by János Körner.)
Elias \cite{Elias1988} established an initial upper bound, showing that $|\mathcal{C}| \le 2(\frac{3}{2})^n$: on the other hand the current best lower-bound is $|\mc C|\gtrsim \left(\frac{9}{5}\right)^{n/4}$ obtained via a semi-random construction in \cite{korner_marton}.
Numerous results on the extensions of the trifference problem to richer alphabets and under varied hashing requirements have been obtained since, including the seminal result on $k$-separating codes (which are $k$-ary analogues of trifferent codes) in \cite{FredmanK1984-perfect} (for a more detailed discussion please see the introduction in \cite{bhandari_khetan}).
However, it has remained challenging to substantially improve the original bound of Elias.
Subsequent works have managed to refine this bound via a computer search for low-dimensional trifferent codes: \cite{FIOREGP2022} and later \cite{Kurz2023trifferent} leading to the improved upper bound of $0.6937 \lrp{\frac{3}{2}}^n$.
A polynomial factor improvement was later obtained in \cite{bhandari_khetan}: more precisely, the bound was improved to $c n^{-2/5} \lrp{\frac{3}{2}}^n$, where $c$ is an absolute constant.
Some variants of trifferent codes have also been considered: such as linear trifferent codes where we think of $\set{0,1,2}$ as $\F_3$ and $\mc C$ is a linear subspace (\cite{PohoataZ22, BishnoiDGP23, della2024upper, dellafiore2025boundskhashdistancesrates}); and more recently in \cite{bishnoi2025generalized} where the demand is that there exist multiple coordinates where any three different code words are mutually distinct (as opposed to a single coordinate in the usual setting).

This article investigates a natural geometric generalization of the concept of trifference, motivated by the aim of introducing new techniques, and of independent interest in its own right.
Borrowing from \emph{vectorial extensions} of other classical combinatorial notions (such as \emph{vector chromatic number} in graph coloring, and semidefinite relaxations of discrete optimization problems~\cite{Lovasz1979,GoemansWilliamson1995}), we extend the discrete alphabet $\{0,1,2\}$ to a continuous one: the set of unit vectors in $\mathbb{R}^3$, which form the $2$-dimensional sphere $S^2$. Furthermore, the combinatorial notion of ``trifference" is replaced with the geometric condition of pairwise orthogonality. This leads to the following definition.
\vspace{1em}
\begin{definition}[(Vector Trifferent Codes)]
	Let $S^2 = \{v \in \mathbb{R}^3 : \|v\| = 1\}$. Consider an integer $n \ge 1$. We say that three points $x,y$ and $z$, all belonging to $(S^2)^n$, are \emph{vector trifferent} if there is an index $i \in \{1, \dots, n\}$ for which the vectors $x_i, y_i, z_i$ are mutually orthogonal.
	A \textit{vector trifferent code} of block length $n$ is a subset $\mathcal{C} \subseteq (S^2)^n$ such that any three distinct codewords $x, y, z \in \mathcal{C}$ are vector trifferent.
\end{definition}

The primary goal of this work is to determine the largest possible size of a vector trifferent code. This geometric variant presents unique challenges that prevent a straightforward application of methods from the discrete setting. We document these challenges below.

\begin{enumerate}
	\item A natural approach is to reduce the geometric problem to its discrete counterpart. 
		To this end, consider the graph $G$ whose vertices are the points on $S^2$ and we join two points on $S^2$ if the corresponding vectors are orthogonal.
		If the chromatic number of $G$ were $3$ then we could have easily created a discrete trifferent code of size $|\mc C|$: consider any proper coloring of $G$ using the colors $\{0,1,2\}$, say $\chi$, and transform $x\in \mc C$ to $(\chi(x_1),\ldots,\chi(x_n)) \in \{0,1,2\}^n$. It is easy to see that the transformed code satisfies the original definition of trifference discussed above.

		Hence, if the above were true we would have deduced that $|\mc C| \leq 2\lrp{\frac{3}{2}}^n$ (or a stronger bound using \cite{bhandari_khetan}).
		However, perhaps surprisingly, the chromatic number of $G$ is $4$ (see \cite{godsil2012_colouring_sphere} for a proof).
		Therefore, the above method transforms a vector trifferent code to a trifferent code over a quaternary alphabet: in other words, a code over $\{0,1,2,3\}^n$ such that for every three distinct codewords $x,y$ and $z$ there is an $i\in [n]$ such that $x_i, y_i$ and $z_i$ are pairwise distinct.
		However, the best known upper bound for such a code is of the order $2^n$ (following the same argument as in \cite{Elias1988}), which is exponentially worse then Elias' original bound.

		Interestingly, \cite{godsil2012_colouring_sphere} also shows that if we restrict only to the set of the \emph{rational points} of $S^2$ (which is a dense subset of $S^2$) then the graph that we get has chromatic number $3$.
		This leads to another attempt where we transform each coordinate of each codeword to a symbol in $\{0,1,2\}$ using the color of a rational point `close' to $x_i$. However, in order for our transformation to produce a trifferent code we would need to know the chromatic number of the graph where not only orthogonal vectors are neighbors but also `almost' orthogonal vectors.
		Thus, in some sense the obstruction to reducing from the vector setting to the familiar discrete setting is ``infinitesimal in nature." 
		% (\todo{will this line be understood well??}) \todo{I think so. The whole thing does have a feeling of "missing by a hairwidth" and the word "infinitesimal" captures that feeling. If this seems inappropriate then we can reword it.}

	\item %\todo{I feel we should mention this as a rather tangential intersection of the vector trifference, since the obvious upper bound does not help the vector trifference problem and any improvement to this upper bound will be entirely useless.}
		%\todo{Sid: I think this has value since it is actually the direct extension of the pruning technique and also shows the connection with some other geometrical questions.}

		Another natural attempt revolves around the following proof technique for the discrete setting. Consider a trifferent code ${\mc C} \subseteq \set{0,1,2}^n$. Now, let $A = \set{0,1}^n$, then we know that $|{\mc C} \cap A|\leq 2$ as any three strings in $A$ are non-trifferent. Combining this with the observation that trifference is a translationally invariant property (translating by $u\in \set{0,1,2}^n$ gives $\mc C \oplus u = \set{x +_{\F_3} u \mid x\in \mc C}$), we can say that the expected size of the intersection of $A$ with a random translate of $C$ is at most $2$: on the other hand it is $|\mc C|\cdot 2^{-n}$. Hence, we get $|\mc C|\leq 2\cdot (3/2)^n$. This is just a reinterpretation of the classical `pruning' style argument for Elias' bound\footnote{Let $\mathcal C_0=\mathcal C\subseteq\{0,1,2\}^n$ be trifferent.  
        For each $i=1$, pick a least‐frequent symbol $a_i\in\{0,1,2\}$ in coordinate $i$ of $\mathcal C_{i-1}$ and delete all codewords with $a_i$ in that coordinate to form $\mathcal C_i$.  
        Then, $|\mathcal C_i|\;\ge (2/3)\,|\mathcal C_{i-1}|.$
        Recursively apply the above process on the pruned set of codewords: this leads to $|\mathcal C_n|\;\ge\;(2/3)^n|\mathcal C|.$         
        Since no three distinct words in $\mathcal C_n$ can all differ in any fixed coordinate, $|\mathcal C_n|\le2$.  
        Hence, $2\;\ge\;|\mathcal C_n|\;\ge\;(2/3)^n|\mathcal C| 
        \text{ which gives } 
        |\mathcal C|\le2\cdot(3/2)^n.$ }
        but was, in fact, exploited in \cite{bhandari_khetan} to get a polynomial improvement.

		Adopting the  above to the vector trifference question involves finding a large subset $A\subseteq S^2$ such that there do not exist three mutually orthogonal points in $A$. Say the Lebesgue measure of $A$, $\mu(A)$, is $\alpha$. Then, we get an upper bound on the size of a vector trifferent code $\mc C$ of block length $n$ as
		%equation
		\begin{equation*}
			|\mc C| \leq 2\lrp{\frac{1}{\alpha}}^n.
		\end{equation*}
		To see why, note that rotating $\mc C$ by a unitary in $(\R^3)^{\otimes n}$, preserves the vector trifference property of $\mc C$.
		Also, the expected number of codewords from $\mc C$ which land in $A^n$ is $\alpha^n \cdot |\mc C|$.
		But this number cannot exceed $2$ as any three points $x,y,z$ in $A^n$ 
		% \todo{again, calling these as vectors may get confusing Sid:points?} can't be vector trifferent. Hence, we get the upper bound claimed.

		So the question is what can be the largest mass of $A$: the larger the better for our upper bound above.
		This question was studied in \cite{perez_thesis} as a natural variant of Wistenhausen's \emph{double cap conjecture} \cite{wistenhausen}.
		In \cite{perez_thesis} it was found that there is an $A$ such that $0.594 \leq \mu((A))$.
		If we apply this result to upper bound the size of a vector trifferent code we get an answer exponentially worse than $(3/2)^n$ which is the approximate bound for the discrete variant.

		In fact, it is actually easy to put an upper bound on $\alpha$ of $2/3$ (simple random argument). This shows that even in the limit this method will not yield upper bounds on vector trifferent codes beyond $2\cdot(3/2)^n$.
		Thus, the problem of vector trifference intersects with some interesting geometrical problems.
		A recent interesting result towards the higher dimensional generalization of this problem appears in \cite{zakharov}.
\end{enumerate}

% %point
% \begin{point}[Another Attempt.]
% 	\todo{please rewrite}
% 	%\label{point:}
% 	Suppose $A\subseteq S^2$ is such that there do not exist three mutually orthogonal points in $A$.
% 	Say $\mu(A) = \alpha$.
% 	We get an upper bound on the size of a vector trifferent code $\mc C$ of block length $n$ as
% 	%equation
% 	\begin{equation*}
% 		|\mc C| \leq 2\lrp{\frac{1}{\alpha}}^n
% 	\end{equation*}
% 	To see why, note that rotating $\mc C$ by a unitary transform preserves vector trifference property of $\mc C$.
% 	Also, the expected number of codewords from $\mc C$ which land in $A^n$ is $\alpha^n/ |\mc C|$.
% 	But this number cannot exceed $2$ and we get the upper bound claimed.

% 	So the question is what can be the largest mass of $A$.
% 	An immediate upper bound on $\alpha$ is $2/3$ (simple random argument).
% 	If this bound could be improved, one would immediately deduce exponential improvement on the bound of a (vector) trifferent code.
% 	An investigation on the largest possible size of $A$ was done in \todo{Perez} and it was found that there is an $A$ such that $\todo{0.594 < \mu((A))}$.
% 	Thus the problem of vector trifference intersects with some interesting geometrical-combinatorial problems.
% 	\href{https://arxiv.org/pdf/2310.06821}{This} paper seems interesting and is quite sophisticated.
% \end{point}

Despite these challenges, this paper establishes an upper bound for vector trifferent codes that is analogous to the bound in the discrete case. Our main result demonstrates that the size of such a code is similarly constrained, and notably, achieves a better leading constant than the standard pruning argument provides for the discrete problem which leads to Elias' original bound for trifferent codes ($2\cdot(3/2)^n$). In fact, before the result of \cite{bhandari_khetan}, the improvements to Elias' bound relied on searching for optimal trifferent codes of low block lengths.

\begin{theorem}[(Main Result)]
	\label{theorem:main theorem}
	Let $\mathcal{C}$ be a vector trifferent code of block length $n$. Then, as $n \to \infty$,
	\[
		|\mathcal{C}| \le (\sqrt{2} + o(1))\left(\frac{3}{2}\right)^n.
	\]
\end{theorem}

The proof appears in Section~\ref{section:proof_of_main_thm}. As discussed, the proof requires techniques distinct from those used in the discrete setting, reflecting the challenges posed by the geometric formulation. 
Along the way we prove a geometrical lemma, Lemma~\ref{lemma:generalized cute lemma}, which might be of independent interest: it concerns the maximum probability that two independently drawn vectors from the same distribution over $\R^d$, are orthogonal.
\emph{Subsequently, we transfer these techniques to the discrete setting and demonstrate that they yield an alternative—similar in spirit yet distinct in method—for improving Elias’s original bound, as achieved in \cite{bhandari_khetan}.}
Please see Section~\ref{section:pairing trick} for more details.
% Then, we generalize the notion to vector trifference to richer alphabets .
\begin{point}[Proof overview.]
	%\label{point:}
 
	Our proof of Theorem \ref{theorem:main theorem} proceeds by translating the combinatorial ``trifference'' condition into a statement about the geometry of certain tensor-product subspaces. The central idea is to associate to each \emph{pair} of distinct codewords $(x, y) \in \mc C^2$ a structured subspace 
	\[
		E_{x,y} \subset (\mathbb{R}^3)^{\otimes n}
	\]
	that captures the coordinates where $x_i$ and $y_i$ are not orthogonal.
	The argument unfolds in three main steps:
	\begin{enumerate}
		\item \textbf{Pairwise subspace construction and orthogonality constraints.}  
			For each coordinate $i$, we define a $1$- or $2$-dimensional subspace $f(x_i, y_i) \subset \mathbb{R}^3$ depending on whether $x_i$ is orthogonal to $y_i$. The tensor product of these subspaces across all coordinates yields $E_{x,y}$. The vector trifference property ensures that for any two \emph{distinct} pairs $(x, y) \neq (a, b)$, the subspaces $E_{x,y}$ and $E_{a,b}$ are orthogonal. 
            % in a strong sense: projecting $E_{a,b}$ onto $E_{x,y}$ annihilates it. This is the geometric analogue of the ``pairing'' trick used in the discrete setting (See Section \ref{section:pairing trick}).

		\item \textbf{Direct sum decomposition and dimension bound.}  
			The orthogonality property above implies that the collection $\{E_{x,y} : x < y\}$ forms a direct sum inside the ambient space $(\mathbb{R}^3)^{\otimes n}$, whose total dimension is $3^n$. Summing dimensions over all pairs $\{x, y\}$ yields the key inequality
			\[
				\binom{|\mc C|}{2} \cdot \underset{\set{x, y}\in \binom{\mc C}{2}}{\mathbb{E}}\left[\,2^{A(x,y)}\,\right] \ \le\ 3^n,
			\]
			where $A(x,y)$ counts the coordinates where $x_i$ and $y_i$ are \emph{not} orthogonal, and the expectation is over a uniformly random draw of $x$ and $y$ from $\mc C$ without replacement.
		\item \textbf{Lower-bounding the expectation.}  
			The quantity $\mathbb{E}_{\set{x, y} \in \binom{\mc C}{2}}[\,2^{A(x,y)}\,]$ is first approximate by $\mathbb{E}_{\set{x, y} \in \mc C \times \mc C}[\,2^{A(x,y)}\,]$ (so $x$ and $y$ are drawn uniformly at random from $\mc C$ with replacement) and then  expanded over subsets $S \subseteq [n]$ of coordinates. For each $S$, the probability that $x_i \not\perp y_i$ for all $i \in S$ is bounded below in two ways:
			\begin{itemize}
				\item \emph{Trivial bound:} at least $1/|\mc C|$, since equality $x = y$ guarantees the event.
				\item \emph{Geometric bound:} appropriately using Corollary \ref{corollary:cute lemma} we can show that for any given $S\subseteq [n]$, the probability that $x_i\not\perp y_i$ for all $i\in S$ is at least $(1/3)^{|S|}$.
			\end{itemize}
			Taking the maximum of these bounds and summing over all $S$ yields a combinatorial expression that can be controlled using Proposition~\ref{prop:estimating_binomial_sums_q_general}. Optimizing over a cutoff parameter balances the two regimes and leads to the asymptotic bound
			\[
				|\mc C| \ \le\ (\sqrt{2} + o(1)) \left(\frac{3}{2}\right)^n.
			\]
	\end{enumerate}
\end{point}

\subsection*{Vector $k$-separating codes}

     In the discrete setting it is natural to generalize the notion of trifference over alphabets of size larger than $3$. 
    \begin{definition}[($k$-separating Codes)]
        \label{defn:q_k_hash}
        Let $k\geq 2$ be an integer.
        A subset $\mc C \subseteq \set{0,1,2,\ldots,k-1}^n$ is called $k$-separating if for any $k$ distinct codewords in $\mc C$ there exists a coordinate $i\in [n]$ such that the symbols of the codewords in the $i^{th}$ coordinate are all mutually distinct.
    \end{definition}

    Hence, the usual trifferent codes are $3$-separating codes in the above terminology.
    In the same spirit, we can define \emph{vector $k$-separating codes}.
    \begin{definition}[(Vector $k$-separating Codes)]
        \label{defn:q_k_hash}
        Let $k\geq 2$ be an integer and consider $S^{k-1}$ the unit sphere in $\R^{k}$.
        A subset $\mc C \subseteq (S^{k-1})^n$ is called vector $k$-separating if for any $k$ distinct codewords in $\mc C$ there exists a coordinate $i\in [n]$ such that the vectors in the  $i^{th}$ coordinate of the codewords are all mutually orthogonal.
    \end{definition}
    Hence, vector trifferent codes are vector $3$-separating codes.     
    The setting of $k$-separating codes for $k> 3$ has been widely  (see Subsection~\ref{subsec:discussions}). The seminal result of  \cite{FredmanK1984-perfect} states that any such code is of size at most 

\begin{equation*}
    \frac{\log_2 |\mc C|}{n}
    \leq \frac{k!}{k^{k-1}} + o(1) \quad \text{  as $n\to \infty$. }
\end{equation*}

    We would like to establish an analogous result for vector $k$-separating codes. 
    A natural approach is to attempt a reduction to the discrete setting; however, as in the case of vector trifferent codes, this strategy encounters the same fundamental obstacles. 
    In particular, the chromatic number of orthogonality graphs on spheres grows exponentially with the dimension \cite{bucic2025geometric}, which prevents a straightforward reduction. 
    Nevertheless, by adapting the proof technique originally developed for $k$-separating codes to the vectorial setting, and making use of Lemma~\ref{lemma:generalized cute lemma}, we are able to obtain the desired result (see Section~\ref{section:fredman_komlos}).

    % We would like to establish an analogous result for vector $k$-separating codes. 
    % A natural approach is to attempt a reduction to the discrete setting; however, as in the case of vector trifferent codes, this strategy encounters the same fundamental obstacles. But we run into the same problem as in the case of the vector trifferent codes since the chromatic number of the orthogonality graphs of spheres grow exponentially in the dimension \href{https://arxiv.org/pdf/2312.06898}{see this}
    % Nevertheless, by adapting the proof technique originally developed for $(k,k)$-separating codes to the vectorial setting, and making use of Lemma~\ref{lemma:generalized cute lemma}, we are able to obtain the desired result (see Section~\ref{section:fredman_komlos}).

% We would like to show something similar for vector $k$-separating codes. 
% Again, the natural way to approach this question would be by reducing it to the discrete setting. However, we are met with the same roadblocks as in the case of vector trifferent codes.
% However, we are able to adapt the proof technique used for $(k,k)$-separating codes to the vectorial setting by using Lemma~\ref{lemma:cute lemma} to obtain the desire result (see section~\ref{section:fredman_komlos}).
    \begin{theorem}[(Upper Bound for Vector $k$-separating Codes)] 
    	\label{thm:vector_kk_separating_codes}
    	Let $\mc C$ be a vector $k$-separating code of block length $n$.
    	Then,
    	$$
    		\frac{\log_2 |\mc C|}{n}
    		\leq \frac{k!}{k^{k-1}} + o(1)
    	$$
        as $n\to \infty$.    
    \end{theorem}

\begin{point}[Discussion \& Open Questions.]
\label{subsec:discussions}
    
	As mentioned, the motivation for introducing vector trifferent codes is to develop new techniques that are applicable to the original discrete setting and that may, in turn, lead to improvements on the long-standing bounds. Our proof technique proceeds by pairing codewords of a vector trifferent code and showing that the corresponding vector spaces associated with these pairs form a direct sum within a \(3^n\)-dimensional ambient space.
    This approach already yields tangible progress: a similar pairing-based analysis can, in fact, be carried out in the discrete setting, leading to an upper bound of $(3/2)^n$, where the leading constant of $1$. Moreover, by incorporating ideas from~\cite{bhandari_khetan}, this bound can be further refined to $O\left(n^{-1/4}\cdot (3/2)^n\right)$,
    which, although weaker than the best-known bounds in the above work, remains of interest as it highlights an alternative line of reasoning towards the problem. (See Section~\ref{section:pairing trick} for more details.)
    This naturally raises the question:
    \begin{quote}
    \textit{Q1. Can the leading constant (currently $\sqrt{2}$) in the bound for vector trifferent codes be improved, perhaps by adapting the discrete pairing-based proof or by other means, and can such a refinement also yield a polynomial improvement (in the spirit of \cite{bhandari_khetan}) over the current upper bound on the size of vector trifferent codes?}
    \end{quote}
    
    From the complementary direction, the construction of large trifferent codes has also remained a challenging problem. 
    A simple random construction shows that there exist trifferent codes in of size approximately $\left({9/7}\right)^{n/2}$. 
    In~\cite{korner_marton}, a stronger construction was given, producing trifferent codes of size $\gtrsim (9/5)^{n/4}$ via a random concatenation technique applied to an inner code of block length $4$ (the so-called \emph{tetra code}). 
    Since every trifferent code is, by definition, also a vector trifferent code, one might expect that constructing vector trifferent codes should be at least as feasible. 
    This naturally raises the question:
    \begin{quote}
    \emph{Q2. Does the continuous geometry of $S^2$ allow for the construction of vector trifferent codes that are exponentially larger than the best-known discrete ones?}
    \end{quote}
    
    Finally, many refinements of the classical bound for $k$-separating codes are known, some tailored to specific values of $k$ and others applying more generally~\cite{Arikan1994, DalaiGR2020, GuruswamiR2019, CostaD2020}. 
    Further extensions have also been explored in which the hashing or separation requirements are relaxed. 
    In particular, one may require only that for some integer $\ell \geq k$, whenever $\ell$ distinct codewords are considered, there exists a coordinate in which all $k$ symbols appear~\cite{ChakrabortyRRS2006, BhandariR2022_IEEE_TOIT}.
    \begin{quote}
    \emph{Q3. Can we adapt these more sophisticated techniques to the vectorial setting?}
    \end{quote}

	% From the other direction, constructing large trifferent codes has also remained a challenging question. 
	% A random construction yields that there are trifferent codes in $\set{0,1,2}^n$ of size approximately $\sqrt(9/7)^n$.
	% In \cite{korner_marton} a construction of better trifferent codes with size $\gtrsim (9/5)^{n/4}$ using a random concatenation technique on top of an inner code of block length $4$ (called the tetra code).
	% Since every trifferent code is naturally a vector trifferent code, constructing vector trifferent codes should only be easier. This leads to the question:

	% \emph{Does the continuous geometry of $S^2$
	% 	allow for the construction of vector trifferent codes that are exponentially larger than the best-known discrete ones.}

	% Of course, the most ambitious question is to obtain  exponential improvements to the upper bound for vector trifferent codes or trifferent codes.
    % \todo{Below we are using q,k or k,k separating codes...have to smoothen the tranisiton to section fredmna komlos}

\end{point}

\section{Technical Preliminaries}
\label{section:technical_preliminaries}

\begin{point}[Notation.]
    If $\mu$ is a probability measure on a space $X$, then we write $\mu\otimes \mu$ to denote the product measure (of $\mu$ with itself) on $X^2$.
    More generally, for any positive integer $\ell$, we write
    $$
        \underbrace{\mu\otimes \cdots \otimes \mu}_{\ell \text{ times}} = \mu^{\otimes \ell}
    $$
    to denote the $\ell$-fold product of $\mu$ with itself, which is a probability measure on $X^\ell$.
\end{point}

Next, we state a lemma that determines the maximum probability of obtaining mutually orthogonal vectors when $\ell$ vectors are drawn i.i.d.\ from a distribution $\mu$ over the $(d-1)$-dimensional sphere. This lemma plays a crucial role in the proofs of Theorems~\ref{theorem:main theorem} and ~\ref{thm:vector_kk_separating_codes}.
% \todo{referecne error}
\begin{lemma}[(Maximum Probability of Mutual Orthogonality Under IID Draws)]
	\label{lemma:generalized cute lemma}
	Let $d \ge 1$ and $\mu \in \mc{P}(S^{d-1})$.
	Let $2\leq \ell \leq d$ be arbitrary.
	Then
	$$
		\mu^{\otimes \ell}(\{(u_1, \ldots, u_\ell) \in (S^{d-1})^\ell : u_i \perp u_j \text{ for all } i\neq j\}) 
		\le \frac{d!}{(d-\ell)!\cdot  d^\ell}
	$$
\end{lemma}
\begin{prf} %proof
	% For the sake of contradiction, suppose the above probability, say $p = \frac{d-1}{d} + \epsilon$ for some $\epsilon > 0$.
	%equation
	Let
	\begin{equation*}
		p
		=\mu^{\otimes \ell}(\{(u_1, \ldots, u_\ell) \in (S^{d-1})^\ell : u_i \perp u_j \text{ for all } i\neq j\}) 
	\end{equation*}
	Fix $n\geq \ell$ arbitrarily.
	Let $V_1, V_2, \dots, V_n$ be $n$ independent random points drawn from $S^{d-1}$ according to the distribution $\mu$. 
	Define a graph $G$ on $[n]$ as follows: we join $i$ and $j$ with an edge if $V_i \perp V_j$.
	Let $I_1, \ldots, I_{\ell}$ be $\ell$ random points chosen from $[n]$ such that $\set{I_1, \ldots, I_\ell}$ is a uniformly random element of $\binom{[n]}{\ell}$.
	Let the probability that the $\ell$ points $V_{I_1}, \ldots, V_{I_\ell}$ are pairwise orthogonal be $p_n$.
	Then the expected number of $\ell$-cliques in $G$ is $p_n\cdot \binom{n}{\ell}$.
	Fix a choice $v_1, \ldots, v_n$ of elements in $S^{d-1}$ such that the corresponding graph $G$ on $[n]$ has at least $p_n \cdot \binom{n}{\ell}$ $\ell$-cliques.
	
	But note that the graph $G$ if $K_{d+1}$-free since there cannot be more than $d$ pairwise orthogonal vectors in $\R^d$.
	By a generalization of Turán's theorem due to Zykov \cite{Zykov1949} (see \href{https://en.wikipedia.org/wiki/Tur%C3%A1n%27s_theorem#Maximizing_Other_Quantities}{Turán's Theorem for Maximizing Other Quantities}), we know that the number of $\ell$-cliques in $G$ is at most 
	%equation
	\begin{equation*}
		\binom{n}{\ell} \cdot \frac{d!}{(d - \ell)! \cdot d^\ell}
	\end{equation*}
	Therefore
	%equation
	\begin{equation*}
		\left\lceil p_n \binom{n}{\ell} \right\rceil
		\leq \binom{n}{\ell}\cdot\frac{d!}{(d - \ell)! \cdot d^\ell}
		\qqra
		p_n + o(1) \leq \frac{d!}{(d - \ell)! \cdot d^\ell}
	\end{equation*}
	Finally, since $p_n\to p$ as $n\to \infty$, we have the claim.
\end{prf}

An immediate consequence of the above is the following, which we record separately.

\begin{corollary}
	\label{corollary:cute lemma}
	Let $d \ge 1$ and $\mu \in \mc{P}(S^{d-1})$. Then
	$$
		\mu \otimes \mu (\{(u,v) \in (S^{d-1})^2 : u \perp v\}) \le \frac{d-1}{d}
	$$
\end{corollary}
Below we state a simple proposition concerning the growth of binomial coefficients, which plays an important role in bounding a key quantity that arises in the proof of Theorem~\ref{theorem:main theorem}.

\begin{proposition}
\label{prop:estimating_binomial_sums_q_general}
Let $q \ge 2$ be an integer and let $\alpha > \tfrac{1}{q+1}$. Then there exists a constant $c_\alpha > 0$ (depending only on $\alpha$) such that

\[
\sum_{m=0}^{\alpha n} \binom{n}{m}\, q^{-m}
\;\ge\;
\left(\frac{q+1}{q}\right)^n \bigl(1 - \exp(-c_\alpha n)\bigr)
\]

for all sufficiently large $n$.
\end{proposition}

\begin{proof}
Let $p = \tfrac{1}{q+1}$ and write $\alpha = p + \varepsilon$ with $\varepsilon > 0$.  
Let $X_1,\dots,X_n$ be independent Bernoulli random variables with $X_i \sim \mathrm{Be}(p)$, and set $X = \sum_{i=1}^n X_i$. Then $\mathbb{E}[X] = np$.  
By the Chernoff bound,
\[
\Pr\!\bigl[X > n\alpha\bigr]
= \Pr\!\bigl[X > n(p+\varepsilon)\bigr]
\;\le\; \exp(-2n\varepsilon^2),
\]
and hence
\[
\Pr\!\bigl[X \le n\alpha\bigr]
\;\ge\; 1 - \exp(-2n\varepsilon^2).
\]
On the other hand,
\[
\Pr\!\bigl[X \le n\alpha\bigr]
= \sum_{m=0}^{\alpha n} \Pr[X = m]
= \sum_{m=0}^{\alpha n} \binom{n}{m} p^m (1-p)^{n-m}.
\]
Rewriting,
\[
\Pr[X \le n\alpha]
= \left(\frac{q}{q+1}\right)^n \sum_{m=0}^{\alpha n} \binom{n}{m} q^{-m}.
\]
Combining the inequalities gives
\[
\sum_{m=0}^{\alpha n} \binom{n}{m}\, q^{-m}
\;\ge\;
\left(\frac{q+1}{q}\right)^n \Bigl(1 - \exp(-2n\varepsilon^2)\Bigr).
\]
Setting $c_\alpha := 2(\alpha - \tfrac{1}{q+1})^2$ completes the proof.
\end{proof}

\section{Proof of Theorem \ref{theorem:main theorem}}
\label{section:proof_of_main_thm}
\begin{proof}[Proof of Theorem~\ref{theorem:main theorem}]
We begin by defining the key local construction that encodes the relationship between two vectors into a subspace of $\mathbb{R}^3$.
Let $\mc S(\R^3)$ denote the set of all the linear subspaces of $\R^3$ and
\[
	f : \mathbb{R}^3 \setminus \{0\} \times \mathbb{R}^3 \setminus \{0\} \ \to\ \mathcal{S}(\mathbb{R}^3)
\]
be given by
\[
	f(u,v) =
	\begin{cases}
		u^\perp \cap v^{\perp}, & \text{if } u \perp v, \\
		u^\perp, & \text{otherwise}.
	\end{cases}
\]
Here $u^\perp$ denotes the orthogonal complement of $u$ in $\mathbb{R}^3$.
For $x,y \in C$, define the \emph{pair subspace}
\[
	E_{x,y} \ :=\ \bigotimes_{i=1}^n f(x_i, y_i) \ \subset\ \bigotimes_{i=1}^n \mathbb{R}^3,
\]
where $\otimes$ denotes the tensor product of spaces.
By construction,
\[
	\dim E_{x,y} \ =\ 2^{A(x,y)},
\]
where
\[
	A(x,y) \ :=\ \big|\{\, i \in [n] : \langle x_i, y_i \rangle \neq 0 \,\}\big|
\]
is the number of coordinates where $x_i$ and $y_i$ are \emph{not} orthogonal.
For any subspace $S \subset \mathbb{R}^3$, let $P_S : \mathbb{R}^3 \to \mathbb{R}^3$ denote the orthogonal projection onto $S$.  
For $x,y \in C$, define
\[
	P_{x,y} \ :=\ \bigotimes_{i=1}^n P_{\,f(x_i, y_i)}.
\]
We now equip $C$ with a fixed total order.
\begin{proposition}
	Let $x,y,a,b \in C$ with $x < y$ and $a < b$ in the total order. Then
	\[
		P_{x,y}(E_{a,b}) \ =\ 
		\begin{cases}
			\{0\}, & \text{if } (a,b) \neq (x,y),\\
			E_{a,b}, & \text{if } (a,b) = (x,y).
		\end{cases}
	\]
\end{proposition}
\begin{proof}
	We consider cases.
	\begin{enumerate}[\quad \emph{Case} 1:]
		\item $a \neq x$ and $a \neq y$.

			In this case, $a, x, y$ are pairwise distinct. By the vector trifference property, there exists $k \in [n]$ such that $a_k, x_k, y_k$ are mutually orthogonal.  
			In this coordinate, $f(x_k, y_k) = x_k^\perp \cap y_k\perp = \mathrm{span}(a_k)$, while $f(a_k, b_k) \subset a_k^\perp = \mathrm{span}\{x_k, y_k\}$.  
			Thus $P_{f(x_k, y_k)}(f(a_k, b_k)) = \{0\}$, forcing $P_{x,y}(E_{a,b}) = \{0\}$.

		\item $a \neq x$ but $a = y$.

			Here $x, a, b$ are pairwise distinct. Again, vector trifference gives a coordinate $k$ where $x_k, a_k, b_k$ are mutually orthogonal.  
			Here $f(a_k, b_k) = a_k^\perp \cap b_k^\perp$, and the projection is onto $f(x_k, y_k) = \mathrm{span}(a_k)$, so the projection is zero.

		\item $a = x$ but $b \neq y$.

			A symmetric argument to Case~2 applies.
	\end{enumerate}

	In all cases except $(a,b) = (x,y)$, the projection is zero.  
	If $(a,b) = (x,y)$, the projection is the identity on $E_{x,y}$ by definition.
\end{proof}

\begin{proposition}
	\[
		\sum_{\substack{x,y \in C\\ x \le y}} E_{x,y} \ =\ \bigoplus_{\substack{x,y \in C\\ x \le y}} E_{x,y}.
	\]
\end{proposition}
\begin{proof}
	Suppose $\sum_{x<y} v_{x,y} = 0$ with each $v_{x,y} \in E_{x,y}$.  
	Applying $P_{x,y}$ and using the previous claim isolates $v_{x,y}$, forcing $v_{x,y} = 0$ for all pairs.  
	Thus the sum is direct.
\end{proof}

\noindent
From this direct sum property, the sum of the dimensions is bounded by the ambient dimension:
\[
	\sum_{\{x,y\} \in \binom{C}{2}} \dim(E_{x,y}) \ \le\ \dim\big((\mathbb{R}^3)^{\otimes n}\big) \ =\ 3^n.
\]
Since $\dim(E_{x,y}) = 2^{A(x,y)}$, we have
\begin{equation}
	\binom{|C|}{2} \cdot \mathbb{E}_{\{x,y\}}[\,2^{A(x,y)}\,] \ \le\ 3^n.
	\tag{1}
\end{equation}
Next, note that
\[
	\underset{(x,y) \in C^2}{\mathbb{E}}[\,2^{A(x,y)}\,]
	= \Pr[x \neq y] \cdot \underset{\set{x, y} \in \binom{C}{2}}{\mathbb{E}}[\,2^{A(x,y)}\,] 
	\quad +\quad  \Pr[x = y] \cdot \underset{x\in C}{\mathbb{E}}[\,2^{A(x,x)}\,].
\]
Since $A(x,x) = n$, this becomes
\[
	\underset{(x,y) \in C^2}{\mathbb{E}}[\,2^{A(x,y)}\,]
	= \frac{|C|-1}{|C|} \cdot \underset{\set{x,y} \in \binom{C}{2}}{\mathbb{E}}[\,2^{A(x,y)}\,] 
	+ \frac{2^n}{|C|}.
\]
Using (1), we deduce
\begin{equation}
	\underset{(x,y) \in C^2}{\mathbb{E}}[\,2^{A(x,y)}\,]
	\ \le\ \frac{2 \cdot 3^n}{|C|^2} + \frac{2^n}{|C|}.
	\tag{2}
\end{equation}
Now expand $2^{A(x,y)}$ as
\[
	2^{A(x,y)} \ =\ \prod_{k=1}^n \big(1 + {1}[\,x_k \not\perp y_k\,]\big)
	\ =\ \sum_{S \subset [n]} \ \prod_{s \in S} {1}[\,x_s \not\perp y_s\,].
\]
Taking expectations,
\[
	\underset{(x,y) \in C^2}{\mathbb{E}}[\,2^{A(x,y)}\,]
	= \sum_{S \subset [n]} \Pr_{(x,y) \in C^2}\big[\,x_s \not\perp y_s \ \forall s \in S\,\big].
\]
For any $S \subset [n]$, two lower bounds hold:
\begin{enumerate}[$\bullet$]
	\item  \emph{Trivial bound:} If $x = y$, the event holds automatically, so
		\[
			\Pr[\,x_s \not\perp y_s \ \forall s \in S\,] \ \ge\ \frac{1}{|C|}.
			\tag{3}
		\]
	\item \emph{Geometric bound:} By Corollary \ref{corollary:cute lemma}, two independent vectors in $\mathbb{R}^3$ are non‑orthogonal with probability at least $1/3$, hence 
		%equation
		\begin{equation*}
			\Pr[\,x_s \not\perp y_s \ \forall s \in S\,] \ \ge\ \left(\frac{1}{3}\right)^{|S|}
			\tag{4}
		\end{equation*}
		This requires some justification.
		We cannot apply Corollary \ref{corollary:cute lemma} by ``successively conditioning," since the independence clause needed by the lemma may not longer be satisfied as soon as we condition even on a single $s\in S$.
		But, fortunately, we can get around this by passing to a suitable space, namely $\bigotimes_{s\in S} \R^3$.
		We equip $\bigotimes_{s\in S}\R^3$ with the natural inner product obtained by tensoring the usual innder product on $\R^3$. 
		The crucial, yet simple observation is
		%equation
		\begin{equation*}
			\Pr[\,x_s \not\perp y_s \ \forall s \in S\,]
			= \Pr\lrb{\otimes_{s\in S} x_s \not \perp \otimes_{s\in S} y_s}
		\end{equation*}
		The vectors $\otimes_{s\in S}x_s$ and $\otimes_{s\in S} y_s$ are unit vectors in $\bigotimes_{s\in S} \R^3$.
		Also, if $x$ and $y$ were picked independently from $C$, then $\otimes_{s\in S}x_s$ and $\otimes_{s\in S}y_s$ are independently picked from the unit sphere in $\bigotimes_{s\in S} \R^3$, both affording the same distribution.
		Thus we can apply Corollary \ref{corollary:cute lemma} and get 
		%equation
		\begin{equation*}
			\Pr[\,x_s \not\perp y_s \ \forall s \in S\,]
			\,=\, \Pr\lrb{\otimes_{s\in S} x_s \not \perp \otimes_{s\in S} y_s}
			\,\geq\, \frac{1}{3^{|S|}}
		\end{equation*}
\end{enumerate}
Combining (3) and (4), we obtain for every $S \subset [n]$:
\begin{equation}
	\Pr_{(x,y) \in C^2}\big[\,x_s \not\perp y_s \ \forall s \in S\,\big]
	\ \ge\ \max\left\{ \frac{1}{3^{|S|}},\ \frac{1}{|C|} \right\}.
	\tag{5}
\end{equation}
Substituting (5) into the expansion of $\mathbb{E}[2^{A(x,y)}]$ gives
\[
	\underset{(x,y) \in C^2}{\mathbb{E}}[\,2^{A(x,y)}\,]
	\ \ge\ \sum_{S \subset [n]} \max\left\{ \frac{1}{3^{|S|}},\ \frac{1}{|C|} \right\}.
\]
Grouping terms by $k = |S|$ and introducing a cutoff parameter $0 < \alpha \le 1$, we split the sum into two regimes:
\begin{equation}
	\mathbb{E}_{(x,y) \in C^2}[\,2^{A(x,y)}\,]
	\ \ge\ \sum_{0 \le k \le \alpha n} \binom{n}{k} \frac{1}{3^k}
	\ +\ \sum_{\alpha n < k \le n} \binom{n}{k} \frac{1}{|C|}.
	\tag{6}
\end{equation}
Recall from (2) that
\begin{equation}
	\underset{(x,y) \in C^2}{\mathbb{E}}[\,2^{A(x,y)}\,] - \frac{2^n}{|C|}
	\ \le\ \frac{2 \cdot 3^n}{|C|^2}.
	\tag{7}
\end{equation}
Combining (6) and (7) yields
\[
	\left[ \sum_{0 \le k \le \alpha n} \binom{n}{k} \frac{1}{3^k}
		\ +\ \frac{1}{|C|} \sum_{\alpha n < k \le n} \binom{n}{k} \right]
	- \frac{2^n}{|C|}
	\ \le\ \frac{2 \cdot 3^n}{|C|^2}.
\]
Rearranging the middle term by noting that
\[
	\sum_{\alpha n < k \le n} \binom{n}{k}
	= 2^n - \sum_{0 \le k \le \alpha n} \binom{n}{k},
\]
we can rewrite the inequality as
\[
	\sum_{0 \le k \le \alpha n} \binom{n}{k} \frac{1}{3^k}
	\ -\ \frac{1}{|C|} \sum_{0 \le k \le \alpha n} \binom{n}{k}
	\ \le\ \frac{2 \cdot 3^n}{|C|^2}.
\]
At this point we invoke Proposition \ref{prop:estimating_binomial_sums_q_general}: for any $\alpha > \tfrac14$, there exists $c_\alpha > 0$ such that
\[
	\sum_{0 \le k \le \alpha n} \binom{n}{k} 3^{-k}
	\ \ge\ \left(\frac{4}{3}\right)^n \left( 1 - e^{-c_\alpha n} \right).
\]
Moreover, for such $\alpha$ the hypergeometric sum $\sum_{0 \le k \le \alpha n} \binom{n}{k}$ is $o(2^n)$, so the second term
\[
	\frac{1}{|C|} \sum_{0 \le k \le \alpha n} \binom{n}{k}
\]
is negligible compared to the first when $|C|$ is exponential in $n$.
Thus, as $n\to \infty$,
\[
	\left(\frac{4}{3}\right)^n (1 - o(1))
	\ \le\ \frac{2 \cdot 3^n}{|C|^2}.
\]
Rearranging gives
\[
	|C|^2 \ \le\ 2 \cdot \left(\frac{3}{2}\right)^n (1 + o(1)),
\]
and hence
\[
	|C| \ \le\ (\sqrt{2} + o(1)) \left(\frac{3}{2}\right)^n.
\]
as $n\to \infty$.
This completes the proof.
\end{proof}

\section{Proof of Theorem ~\ref{thm:vector_kk_separating_codes}}
\label{section:fredman_komlos}

We now turn our attention to vector $k$-separating codes as mentioned in subsection~\ref{subsec:discussions}. 

\begin{proof}[Proof of Theorem~\ref{thm:vector_kk_separating_codes}]
	Let $\mc C\subseteq (S^{k-1})^n$ be a vector $k$-separating code of block length $n$.
	We want to put an upper bound on the size of $\mc C$.
	Let $\ell = k-2$ and $X_1, \ldots, X_\ell$ be randomly chosen from $\mc C$ such that the set $\mc X = \set{X_1, \ldots, X_\ell}$ is a uniformly random element of $\binom{\mc C}{\ell}$.
	Consider the complete graph $G_{\mc X}$ whose vertex-set is $\mc C \setminus \mc X$.
	For distinct vertices $a$ and  $b$ of $G_{\mc X}$, we color the edge between $a$ and $b$ by the symbol $j$ if $j$ is the smallest element in $[n]$ such that $a_j, b_j, X_1, \ldots, X_\ell$ are pairwise orthogonal.
    Such a $j$ must exist because of the $k$-separating property of the code $\mc C$.
    
	Let $G^{(j)}_{\mc X}$ denote the graph whose vertex-set is $\mc C \setminus \mc X$ and whose edge set consists of all the $j$-colored edges.
	Note that $G_{\mc X}^{(j)}$ is a bipartite graph.
    This is because, if there is at least one $j$-colored edge, then the vectors $X_1(j), \ldots, X_\ell(j)$ span a $\ell(=k-2)$-dimensional subspace of $\R^k$, and the $j$-th entries of the endpoints any $j$-colored edge lie in the orthogonal complement of this subspace, which is just a copy of $\R^2$.
    It follows that any odd cycle in $G_{\mc X}^{(j)}$ produces an odd-length closed walk in the orthogonality graph of $S^1$.
    But this latter graph is bipartite since it is a disjoint union of $4$-cycles.
    Therefore, there cannot be any odd cycle in $G_{\mc X}^{(j)}$ and hence it is a bipartite graph.
    
    It is clear that
	%equation
	\begin{equation*}
		G_{\mc X}
		= G_{\mc X}^{(1)} \cup \cdots \cup G_{\mc X}^{(n)}
	\end{equation*}
	If $\tau_{\mc X}^{(j)}$ denote the fraction of non-isolated vertices of $G_{\mc X}^{(j)}$, then by Hansel lemma we know that
	%equation
	\begin{equation*}
		\log(|\mc C| - \ell) = \log|G_{\mc X}| \leq \sum_{j=1}^n \tau_{\mc X}^{(j)}
	\end{equation*}
	Taking expectation, we can write
	%equation
	\begin{equation*}
		\log(|\mc C| - \ell)
		\leq \sum_{j=1}^n \E[\tau_{\mc X}^{(j)}]
		\tag{*}
	\end{equation*}
	Let us think about how to upper bound $\E[\tau_{\mc X}^{(j)}]$.
	This is same as the probability that a vertex in $G_{\mc X}$ selected uniformly at random is non-isolated in $G_{\mc X}^{(j)}$.
	This is clearly upper bounded by the probability that a vertex $V$ in $G_{\mc X}^{(j)}$ selected uniformly at random is such that the vectors $X_1, \ldots, X_{\ell}, V$ are pairwise orthogonal.
	But $\set{X_1, \ldots, X_\ell, V}$ is just a uniformly random subset of $\binom{\mc C}{\ell + 1}$.
	For the sake of more convenient notation, let $\mc S = \set{S_1, \ldots, S_{\ell + 1}}$ be a uniformly random subset of $\mc C$ of size $\ell + 1$.
	Thus, form (*), we have
	%equation
	\begin{equation*}
		\log(|\mc C| - \ell) \leq n \Pr[S_i \perp S_j \text{ for all } i \neq j]
		\quad \Rightarrow\quad
		\frac{\log(|\mc C| - \ell)}{n}
		\leq \Pr[S_i \perp S_j \text{ for all } i \neq j]
		\tag{**}
	\end{equation*}
    We would like to invoke Lemma \ref{lemma:generalized cute lemma} to upper bound the RHS of the above to finish the proof.
    But to do that we need to work with samples \emph{with} replacement.
    To this end, let $Y_1, \ldots, Y_{\ell + 1}$ be uniformly and independently chosen from $\mc C$.
	Noting that
    \begin{equation*}
    \begin{aligned}
        &\Pr[Y_i\perp Y_j \text{ for all } i\neq j]\\
        &= \Pr[Y_i\perp Y_j \text{ for all } i\neq j|\ Y_1, \ldots, Y_{\ell + 1} \text{ are pairwise distinct}] \cdot \Pr[Y_1, \ldots, Y_{\ell + 1} \text{ are pairwise distinct}]\\
        &= \Pr[S_i\perp S_j \text{ for all } i\neq j] \cdot
        \lrb{
            \lrp{1 - \frac{1}{|\mc C|}} \cdot \lrp{1 - \frac{2}{|\mc  C|}} \cdot
            \cdots
            \cdot
            \lrp{1 - \frac{\ell}{|\mc C|}}
        }\\
        &\geq \Pr[S_i\perp S_j \text{ for all } i\neq j] \cdot \lrp{1 - \frac{\ell}{|\mc C|}}^\ell
    \end{aligned}
    \end{equation*}
    
    From this and (**) we get

	\begin{equation*}
    \begin{aligned}
		\frac{\log(|\mc C| - \ell)}{n}
		\leq \Pr[Y_i \perp Y_j \text{ for all } i \neq j] \cdot \frac{1}{\lrp{1 - \ds\frac{\ell}{|\mc C|}}^\ell}
        \leq \Pr[Y_i\perp Y_j \text{ for all } i \neq j] \cdot \frac{1}{1 - \ds\frac{\ell^2}{|\mc C|}}
    \end{aligned}
	\end{equation*}
    where we have used the fact that $(1 - x)^n \geq 1 - nx$ if $x \in (0, 1)$ and $n\geq 1$.
   This gives
    \begin{equation*}
        \lrp{1 - \ds\frac{\ell^2}{|\mc C|}} \cdot \frac{\log(|\mc C| - \ell)}{n} \leq
        \Pr[Y_i\perp Y_j \text{ for all } i \neq j].    \tag{\textdagger}   
    \end{equation*}
    
	But by Lemma \ref{lemma:generalized cute lemma} we also know that
	%equation
	\begin{equation*}
		\Pr[Y_i \perp Y_j \text{ for all } i \neq j]
		\leq \frac{k!}{(k - (\ell + 1))!\cdot k^{\ell + 1}}
		= \frac{k!}{k^{k-1}}
	\end{equation*}
    Using this in ($\dagger$) we get
    \[
         \lrp{1 - \ds\frac{\ell^2}{|\mc C|}} \cdot \frac{\log(|\mc C| - \ell)}{n} \leq
         \frac{k!}{k^{k-1}}
    \]
	The desired result is now obvious since $|\mc C|$ is large compared to $\ell$.
\end{proof}

\section{Reanalyzing the Discrete Setting Using the Proof of Theorem~\ref{theorem:main theorem}}
\label{section:pairing trick}
We now focus our attention to classical discrete setting.
Let $\Sigma = \set{0, 1, 2}^n$ and $\mc C \subseteq \Sigma^n$ be a trifferent code.

\begin{enumerate}
    \item First, we show how the proof idea of Theorem~\ref{theorem:main theorem} shows that $|\mc C|<(1+o(1))\cdot (3/2)^n $
    \item Then, we show how we can import an idea from \cite{bhandari_khetan} to show $|\mc C|\lesssim (n^{-1/4}\cdot (3/2)^n)$.
\end{enumerate}

\begin{proposition}
\label{prop:pairing_easy}
    $|\mc C|\leq (1+o(1))\cdot (3/2)^n$.
\end{proposition}

\begin{proof}

Define a bipartite graph $G = (U, V)$, where
$$
U = \binom{\mathcal C}{2}
\quad\text{ and }\quad
V = \Sigma^n,
$$
and we join an edge between $\set{x, y}\in U$ with $a\in V$ if and only if both $x$ and $y$ differ from $a$ in every coordinate.
More precisely, we join an edge between $\set{x, y}\in U$ and $a\in V$ if
$$
x_i \neq a_i
\quad \text{and} \quad
y_i \neq a_i
$$
for all $i\in [n]$.
The fact that $\mathcal C$ is a trifferent code implies that no vertex in $V$ can have degree greater than $1$.
Also, for each $\set{x, y}$ in $U$ we have
$$
\deg(\set{x, y}) = 2^{A(x, y)}
$$
where $A(x, y) = |\set{i\in [n]:\ x_i = y_i}|$.
By double counting, we have
%equation
\begin{equation*}
	\sum_{u\in U} \deg(u) = \sum_{v\in V} \deg(v)
\end{equation*}
We will lower bound the left hand side and upper bound the right hand side to obtain the desired bound.
The left hand side is
%equation
\begin{equation*}
	\sum_{u\in U} \deg(u)
	= \sum_{\set{x, y} \in \binom{\mc C}{2}} \deg(\set{x, y})
	= \sum_{\set{x, y}\in \binom{\mc C}{2}} 2^{A(x, y)}
	= \binom{|\mc C|}{2} \underset{\set{x, y}\in \binom{\mc C}{2}} \E[2^{A(x, y)}]
	\tag{*}
\end{equation*}

%lemma
Next, we show the following
	%\label{lemma:}
	$$
		\lrp{\frac{4}{3}}^n (1 - o(1))
		\leq \underset{\set{x, y}\in \binom{\mc C}{2}}{\E}[2^{A(x, y)}].
	$$

 %proof
	The notion of randomness we have chooses a $2$-element subset of $\mc C$ uniformly at random.
	It is convenient to convert this random model to the one where we choose two elements of $\mc C$ successively, with repetition allowed.
	To this end, we write
	%equation
	\begin{equation*}
		\underset{(x, y)\in \mc C^2}\E[2^{A(x, y)}]
		=
		\underset{(x, y) \in \mc C^2}\Prob[x\neq y]\cdot \underset{\set{x, y} \in \binom{\mc C}{2}}\E[2^{A(x, y)}] \quad + \underset{(x, y) \in \mc C^2}\Prob[x = y] \cdot 2^n
		%\tag{*}
	\end{equation*}
	Thus
	%equation
	\begin{equation*}
		\underset{(x, y) \in \mc C^2}\E[2^{A(x, y)}]
		= \lrp{1 - \frac{1}{|\mc C|}} \underset{\set{x, y} \in \binom{\mc C}{2}}{\E}[2^{A(x, y)}] + \frac{2^n}{|\mc C|}
		\tag{1}
	\end{equation*}
	% Hence
	% %equation
	% \begin{equation*}
	% 	\frac{|\mc C|}{|\mc C| - 1} \underset{(x, y) \in \mc C^2}{\E}[2^{A(x, y)}] - \frac{2^n}{|\mc C| - 1} 
	% 	= \underset{\set{x, y} \in \binom{\mc C}{2}}{\E}[2^{A(x, y)}]
	% 	\tag{2}
	% \end{equation*}
	% and hence
	% %equation
	% \begin{equation*}
	% 	\frac{|\mc C|^2}{2} \underset{(x, y)\in \mc C^2}{\E}[2^{A(x, y)}] - \frac{|\mc C|}{2}\cdot 2^n
	% 	= \binom{|\mc C|}{2} \underset{\set{x, y}\in \binom{\mc C}{2}}{\E}[2^{A(x, y)}]
	% \end{equation*}
	So we will lower bound the left hand side in the above.
	We have
	%equation
	\begin{equation*} 
		2^{A(x, y)}
		= \prod_{i\in [n]} (1 + 1[x_i = y_i])
		= \sum_{S\subseteq [n]} 1[x_S = y_S]
	\end{equation*}
	where ``$x_S = y_S$" means that $x_s = y_s$ for all $s\in S$.
	Since the probability of $x_S = y_S$ is at least the collision probability for two uniformly and independently chosen elements from $\Sigma^S$, we have
	%equation
	\begin{equation}
		\underset{(x, y) \in \mc C^2}\E[1[x_S = y_S]]
		\geq (1/3)^{|S|}
	\end{equation}
	We also have the \emph{trivial bound} that 
		\[
			\Pr[\,x_S =  y_S\,]
			\geq \Pr[x = y]
			\geq \frac{1}{|\mc C|}.
		\]
	Combining these, we obtain for every $S \subset [n]$:
	\begin{equation*}
		\Pr_{(x,y) \in C^2}\big[\,x_S = y_S\,\big]
		\geq \max\left\{ \frac{1}{3^{|S|}},\ \frac{1}{|\mc C|} \right\}
		\qqra
		\underset{(x,y) \in C^2}{\mathbb{E}}[\,2^{A(x,y)}\,]
		\geq \sum_{S \subset [n]} \max\left\{ \frac{1}{3^{|S|}},\ \frac{1}{|\mc C|} \right\}.
	\end{equation*}
	Grouping terms by $k = |S|$ and introducing a cutoff parameter $0 < \alpha \le 1$, we split the sum into two regimes:
	\begin{equation}
		\mathbb{E}_{(x,y) \in C^2}[\,2^{A(x,y)}\,]
		\ \ge\ \sum_{0 \le k \le \alpha n} \binom{n}{k} \frac{1}{3^k}
		\ +\ \sum_{\alpha n < k \le n} \binom{n}{k} \frac{1}{|C|}.
		\tag{2}
	\end{equation}
	Using this in (1) we get
	%equation
	\begin{equation*}
		\sum_{0 \le k \le \alpha n} \binom{n}{k} \frac{1}{3^k} + \sum_{\alpha n < k \le n} \binom{n}{k} \frac{1}{|C|}
		\leq \lrp{1 - \frac{1}{|\mc C|}} \underset{\set{x, y} \in \binom{\mc C}{2}}{\E}[2^{A(x, y)}] + \frac{2^n}{|\mc C|}
	\end{equation*}
	which gives
	%equation
	\begin{equation*}
		\sum_{0 \le k \le \alpha n} \binom{n}{k} \frac{1}{3^k} + \lrb{\sum_{\alpha n < k \le n} \binom{n}{k} \frac{1}{|\mc C|} - \frac{2^n}{|\mc C|}}
		\leq \lrp{1 - \frac{1}{|\mc C|}} \underset{\set{x, y} \in \binom{\mc C}{2}}{\E}[2^{A(x, y)}]
	\end{equation*}
	Writing $2^n$ as $\sum_{k=0}^n \binom{n}{k}$ we get
	\begin{equation*}
		\sum_{0 \le k \le \alpha n} \binom{n}{k} \frac{1}{3^k} + \sum_{0 \leq k\leq \alpha n} \binom{n}{k} \frac{1}{|\mc C|}
		\leq \lrp{1 - \frac{1}{|\mc C|}} \underset{\set{x, y} \in \binom{\mc C}{2}}{\E}[2^{A(x, y)}]
	\end{equation*}
	The first term in the left hand side is $\lrp{\frac{4}{3}}^n (1 - o(1))$ (see Proposition~\ref{prop:estimating_binomial_sums_q_general}) and the second term is $o(2^n) / |\mc C|$ whence the lemma follows.

From the above argument and from (*) we have
%equation
\begin{equation*}
	\binom{|\mc C|}{2} \lrp{\frac{4}{3}}^n (1 - o(1)) \leq \sum_{u\in U} \deg(u)
	\tag{**}
\end{equation*}
Now for the right hand side.
We have 
%equation
\begin{equation*}
	\sum_{v\in V}\deg(v)
	= \sum_{x\in \Sigma^n} \deg(x)
	= 3^n \underset{x\in \Sigma^n}{\E}[\deg(x)]
\end{equation*}
But note that, by definition of our bipartite graph and by the fact that $\mc C$ is a trifferent code, we have $\deg(v)$ is either $0$ or $1$ for any $v\in V$.
In fact, if for any $v\in V$ we define
%equation
\begin{equation*}
	H_v = \set{x\in \Sigma^n: \Delta(x, v) = n}
	\quad \text{ and } \quad 
	h_v = |H_v\cap \mc C|
\end{equation*}
then $\deg(v) = \binom{h_v}{2}$.
Thus
%equation
\begin{equation*}
		\sum_{v\in V} \deg(v)
        = 3^n \cdot \Pr_v[h_v = 2]
		= 3^n\cdot \E_v [1[h_v = 2]]
\end{equation*}
Applying Markov, we get
\begin{equation*}
	\begin{aligned}
        \sum_{v\in V} \deg(v)
		&\leq 3^n\cdot \frac{\E_v[h_v]}{2}\\
		&= \frac{3^n}{2} \cdot\underset{v}{\E}\lrb{\sum_{x\in \mc C} 1[x\in H_v]}\\
		&= \sum_{x\in \mc C} \frac{3^n}{2} \cdot \Pr_v[x\in H_v]\\
		&= \frac{3^n}{2} \cdot |\mc C| \cdot \lrp{\frac{2}{3}}^n = 2^{n-1}\cdot |\mc C|
	\end{aligned}
\end{equation*}
Using this in (**) and using the double counting, we have
%equation
\begin{equation*}
	\binom{|\mc C|}{2}\cdot \lrp{\frac{4}{3}}^n
	\leq 2^{n-1} |\mc C|
	\qqra 
	|\mc C| - 1 \leq \lrp{\frac{3}{2}}^n
\end{equation*}
which shows that $|\mc C| = (1 + o(1))\lrp{\frac{3}{2}}^n$.
\end{proof}

Next, we give a brief proof sketch on adapting the idea in \cite{bhandari_khetan} to obtain a polynomial factor improvement.

\begin{proposition}
\label{prop:pairing_hard}
    $|\mc C| \lesssim n^{-1/4}\cdot (3/2)^n$
\end{proposition}

\begin{proof}[Proof Sketch]
    To show this we adapt the proof of Proposition~\ref{prop:pairing_easy}.
    We modify the definition of the bipartite graph $G = (U, V)$ as below. The modification is to relax the criterion used in the above proof which governs the edge structure.
    Let $Agr(x,y)$ denote the set of coordinates where $x$ and $y$ agree.
    $$
    U = \binom{\mathcal C}{2}
    \quad\text{ and }\quad
    V = \Sigma^n,
    $$
    and we join an edge between $\set{x, y}\in U$ with $a\in V$ iff
    \[
    |\set{i \in Agr(x,y)\mid a_i\in \set{x_i, y_i}}| = 2 \quad \text{ and } \quad |\set{i \notin Agr(x,y)\mid a_i\in \set{x_i, y_i}}| = 0.
    \]
    In the proof above, the condition was $|\set{i \in [n]\mid a_i\in \set{x_i, y_i}}| = 0$.
    Hence, we have relaxed the condition for edge existence whenever $i \in Agr(x,y)$.
    To understand this condition better let us take the example of $a = \mathbf{2}$, i.e., the all $2$'s string.
    By the above condition a pair of strings $\set{x,y}$ is joined to $a$ iff there are at most two coordinates $i\in [n]$ such that the symbol $2$ appears in ${x_i, y_i}$ and in those coordinates $x_i =y_i$.  
    
    The questions now becomes what are the left and right degree of this bipartite graph. 
    \begin{enumerate}
        \item For the left degree: this can be computed in a manner similar to the one used in Proposition~\ref{prop:pairing_easy}. It turns out that there is roughly a quadratic increase in the left degree due to the relaxed conditions, i.e.,
        \[
            n^2\times \binom{|\mc C|}{2} \lrp{\frac{4}{3}}^n \lesssim \sum_{u\in U} \deg(u)
        \]
        \item For the right degree: the fact that $\mathcal C$ is a trifferent code implies that no vertex in $V$ can have degree greater than $\approx n^{3/2}$. 
        To see this consider a string in $V$, say $a = \mathbf{2}$. Let $N(a)$ be the neighborhood of $a$. Construct a graph $H$ where the vertex set is $[n]$ and we join $i$ and $j$ if there exists a pair $\set{x,y} \in N(a)$ such that $x_i = y_i = 2$. Hence, the number of edges in $H$ is $|N(a)|$.
        Following the arguments from \cite{bhandari_khetan}, it is possible to verfiy that $H$ is $4$-cycle free and then by the famous result of K\H{o}v\'{a}ri, S\'{o}s and Tur\'{a}n \cite{KovariSosTuran1954}, known popularly as the KST theorem, we have our desired bound.
    \end{enumerate}
    % The fact that $\mathcal C$ is a trifferent code implies that no vertex in $V$ can have degree greater than $1$.
    Combining the above observations yields the $n^{-1/4}$ improvement.
\end{proof}

\bibliographystyle{alpha}
\bibliography{VT.bib}
\end{document}